\documentclass[draft]{agujournal2019}
\usepackage{url}
\usepackage{lineno}
\usepackage[inline]{trackchanges}
\usepackage{soul}
\usepackage{multirow, makecell}
\usepackage{amsmath}
\usepackage{tabularx}
\usepackage{comment}

\draftfalse

\journalname{JGR: Space Physics}

\begin{document}

\title{Response of Ionospheric Total Electron Content to the Impulsive and Late Phases
of X-Class Solar Flares with Various Center-to-Limb Locations}

\authors{S. Z. Bekker\affil{1}, R. O. Milligan\affil{1}, and I. A. Ryakhovsky\affil{2}}

\affiliation{1}{Queen's University Belfast, Belfast, United Kingdom}
\affiliation{2}{Sadovsky Institute of Geospheres Dynamics, Russian Academy of Sciences, Moscow, Russia}

\correspondingauthor{Susanna Bekker}{s.bekker@qub.ac.uk, susanna.bekker@gmail.com}

\begin{keypoints}
\item An empirical relationship between the $TEC$ increase and the solar flux increase during the impulsive and late flare phases has been obtained
\item The influence of the solar flare's location on the increase in ionospheric $TEC$ during different phases has been demonstrated
\item The latitudinal influence on the increase in electron concentration during the late emission of warm coronal lines has been observed
\end{keypoints}

%
%

%
%


\begin{abstract}

During a solar flare, the fluxes in various lines and continua of the solar spectrum increase, leading to enhanced ionization of the illuminated part of the Earth's ionosphere and an increase in the total electron content ($TEC$). It has been previously shown that nearly 50\% of X-class solar flares exhibit a second peak in warm coronal lines, such as Fe~XV and Fe~XVI, (called the "EUV late phase") the effect of which on the ionosphere remains largely unexplored. This study presents an analysis of the ionospheric response to 14 X-class flares with pronounced late phases. For the first time, empirical relationships between the increase in $TEC$ and the solar flux enhancement during the impulsive and late phases of the flare are derived. Additionally, we demonstrate the influence of flare location on the intensity of geoeffective solar spectral lines and the ratio of the ionospheric responses to the impulsive and late phases of solar flares.
\end{abstract}

\section*{Plain Language Summary}

This paper investigates the effects of different phases of solar flares on the state and dynamics of the Earth's ionosphere. Although the impact of the main phase of solar flares has been extensively studied, the influence of late-phase radiation bursts on the Earth's ionosphere remains largely unexplored. In this work, solar and ionospheric parameters were analyzed during 14 powerful solar flares with pronounced late phases. For the first time, empirical relationships between the ionospheric electron concentration and solar flux were established during the main and late phases of solar flares. Additionally, the study identified a connection between this effect on the ionosphere and the flare location on the solar disk.

\section{Introduction}

The primary sources of ionization in the Earth's ionosphere are extreme ultraviolet (EUV) and $X$-ray radiation from the Sun. The processes occurring in the ionosphere due to solar radiation have been studied for decades, both through experimental measurements and theoretical models. These models now provide a good understanding of the diurnal and seasonal variations in ionospheric parameters, as well as changes driven by the 11-year solar activity cycle. However, more sudden and rapid phenomena, such as solar flares, geomagnetic storms, and traveling ionospheric disturbances, remain less understood. The accuracy of predicting their effects on ionospheric layers is still significantly lower compared to predictions under quiet solar conditions.

During solar flares, there is a significant increase in solar radiation across a broad range of wavelengths, leading to a non-uniform rise in electron and ion concentrations at various altitudes in the illuminated part of the ionosphere \cite{Mitra_1974, Leonovich_etal_2002, Xiong_etal_2011, deAbrau_etal_2019, Bekker_etal_2021}. The magnitude and delay of the ionospheric response to variations in the solar spectrum depend on several factors, including the temporal dynamics of radiation flux, background solar and magnetic activity, season, solar zenith angle, and geographical latitude. It is well-established that $X$-ray radiation during solar flares penetrates the lower ionosphere, ionizing the primary neutral components of the $D$ region (50–90 km) and partially affecting the $E$ region (90–120 km), which is also ionized by Lyman-$\beta$ radiation \cite{Brasseur_Solomon_2005, Xu_etal_2021}. The higher $F1$ and $F2$ layers are primarily ionized by EUV radiation (10–100 nm), which is responsible for the main increase in total electron content ($TEC$) in the ionosphere during flares \cite{Mitra_1974, Nishimoto_etal_2023}.

Different emissions in the EUV range of the solar spectrum contribute differently to the ionization rate. This contribution is influenced by the magnitude and temporal dynamics of the flux in specific lines or continua, as well as by the ionization and absorption cross-sections and the concentration of the particles to be ionized \cite{Solomon_Qian_2005, Bekker_2018, Korsunskaja_2019, Watanabe_etal_2021, Nishimoto_etal_2023}. Considering these factors, the most geoeffective lines of the solar spectrum during the impulsive phase of a solar flare are the cool chromospheric lines He~II 30.4 nm, He~I 58.4 nm, and C~III 97.7 nm, with He~II 30.4 nm being the most geoeffective \cite{Watanabe_etal_2021, Nishimoto_etal_2023, Bekker_etal_2024}. Previously, \citeA{Woods_etal_2011} have shown that a significant portion of solar flares are followed by emissions from warm coronal lines (such as Fe~XV 28.4 nm and Fe~XVI 33.5 nm), which can occur minutes to hours after the impulsive phase. These emissions are known as the EUV late phase of the solar flare and are typically not accompanied by an increase in other parts of the solar spectrum (including the most geoeffective lines of the main phase, such as Fe~XX 13.3 nm, He~II 30.4 nm, He~I 58.4 nm, C~III 97.7 nm, and $X$-rays). Thus, in the absence of other significant ionization sources during the late phase of a solar flare, the effect of warm coronal lines on $TEC$ variations can be clearly observed.

To date, the effect of the late phase of a solar flare on the increase in electron concentration ($Ne$) in the ionosphere has been demonstrated for only two flares: one on November 3, 2011 \cite{Bekker_etal_2024} and another on October 23, 2012 \cite{Liu_etal_2024}. These studies showed that the $TEC$ response to the previously overlooked late warm coronal emissions can be detected in experimental data, and the response is comparable to the well-studied effects of the main phase of a solar flare, which has been extensively studied over the past several decades using data from global navigation satellite systems (GNSS) \cite{Wan_etal_2002, Wan_etal_2005, Meza_etal_2006, Garcia_etal_2007, Guyer_Can_2013, Le_etal_2013, Hazarika_etal_2016, Yasyukevich_etal_2018, Habarulema_etal_2020, Bekker_Ryakhovsky_2024, Sreeraj_etal_2025}, ionosondes \cite{Barta_etal_2019, Habarulema_etal_2022, Buzas_etal_2023}, incoherent scatter radars \cite{Qian_etal_2019, Setov_etal_2020}, very low frequency (VLF) signal receivers \cite{Palit_etal_2013, Kumar_Kumar_2018, Hayes_etal_2021, Bekker_etal_2022, Kolarski_etal_2022, Bekker_Korsunskaya_2023, Ryakhovsky_etal_2024}, and other instruments. Despite the fact that Fe~XV 28.4 nm and Fe~XVI 33.5 nm emissions have high geoeffectiveness and contribute to the increase in charged particles in the ionospheric $F$ region, as of 2024, their effect on the Earth's ionosphere has not been analyzed. There are several reasons for this. First, the contribution of warm coronal lines is often overshadowed by the stronger ionizing radiation from the main (impulsive + gradual) phase of a solar flare. Second, the increase in flux in the Fe~XV 28.4 nm and Fe~XVI 33.5 nm lines during the late phase occurs gradually, making $TEC$ variations less obvious compared to the more sharp changes caused by the impulsive phase. These factors combined have resulted in the limited study of the EUV late phase's effect on the Earth's ionosphere. This work aims to address this gap.

A large volume of GNSS experimental data has been accumulated over several solar activity cycles, providing valuable insights into the behavior of ionospheric parameters. These data enable the investigation of the ionospheric response to the EUV late phase, which accompanies $\sim$50\% of X-class flare \cite{Woods_etal_2011, Woods_Thomas_2014, Liu_etal_2015, Dai_etal_2018, Zhou_etal_2019, Chen_etal_2020, Zhong_etal_2021, Liu_etal_2024, Liu_etal_2024_2}. By analyzing solar observations in the $X$-ray and EUV ranges from instruments aboard the SDO (Solar Dynamics Observatory; \cite{Pesnell_etal_2012}) and GOES (Geostationary Operational Environmental Satellite; \cite{Machol_Viereck_2016}), as well as ionospheric measurements from the SOPAC (Scripps Orbit and Permanent Array Center) GNSS network, we examined the effect of 14 X-class flares with pronounced late phases on the Earth's ionosphere. These flares, which occurred between 2011 and 2024, exhibited similar power levels (from X1.0 to X3.0) but had vastly different spectra, partly due to variations in their locations on the solar disk. Since solar spectral lines are formed in different regions of the solar atmosphere, the measured spectral shape depends on the heliocentric angle \cite{Woods_etal_2005, Chamberlin_etal_2020, Thiemann_etal_2018}. The objective of this work is to use GNSS data to numerically investigate the dynamics of electron concentration in the Earth's ionosphere during X-class flares and to derive an empirical relationship between $TEC$ increase and irradiance increase during both the impulsive and late phases of the flare. This is the first multi-event statistical study to directly correlate warm coronal emission flux with $TEC$ enhancements during the late phase. A key aspect of this study is analyzing the influence of the flare’s location on the solar disk, which affects its spectrum and subsequently influences the ratio between the ionospheric response to the impulsive and late phases of the flare.

Section 2 of this paper examines the center-to-limb variations of cold chromospheric and warm coronal emissions. Section 3 outlines the experimental solar and ionospheric data utilized in this study, along with the methodology for estimating $TEC$ from GNSS data. Section 4 presents the empirical relationship between $TEC$ increase and flux variations during both the impulsive and late phases, while also analyzing the impact of flare location on the ionospheric response. Section 5 provides a discussion of the results.

\section{Center-to-Limb Variations of the $X$-ray and EUV Emissions}

\begin{figure} [h]
        \noindent\includegraphics[width=420pt]{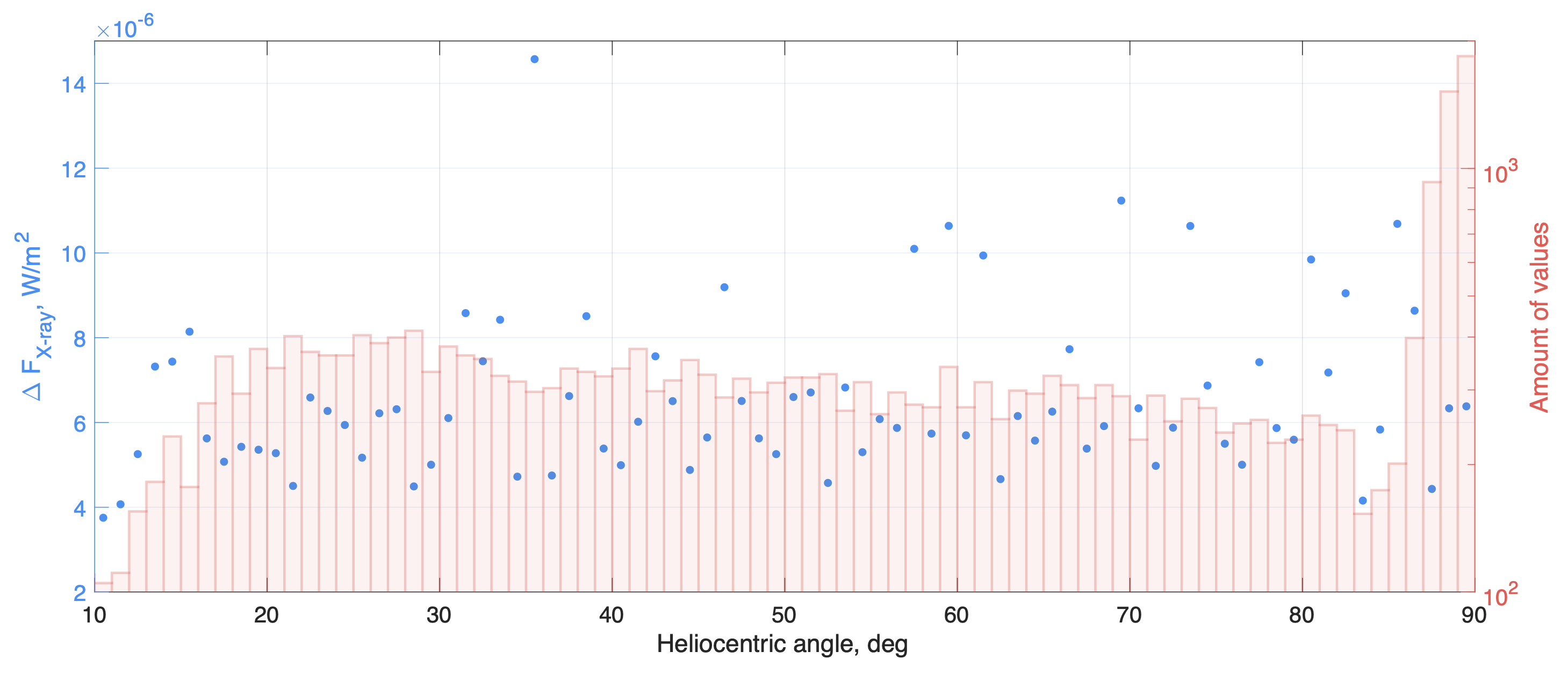}
        \centering
        \caption{Center-to-limb variations of $X$-ray flux (0.1–0.8 nm) during solar flares. The blue dots represent the average values of $\Delta F_{Xray}$ for each degree of heliocentric angle; the red histogram bars show the number of processed values (flares).}
    \label{Center_to_limb_Xray}
    \end{figure}

It is well known that different emissions of the solar spectrum undergo varying levels of attenuation depending on the flare location on the solar disk \cite{Worden_etal_2001, Woods_etal_2006, Milligan_2021}. This variation is primarily due to the region in which the emission is generated. For instance, $X$-ray radiation, which is produced in the solar corona or transition region, is optically thin. As a result, its intensity remains nearly constant, regardless of the flare's position relative to the center of the solar disk. To verify this, we analyzed 16,747 C-, M-, and X-class flares that occurred between October 20, 2006, and June 30, 2024, as cataloged by the Hinode space observatory \cite{Watanabe_etal_2012}. Figure~\ref{Center_to_limb_Xray} displays the change in intensity of $X$-ray radiation in the 0.1–0.8 nm range ($\Delta F_{Xray}$) during these flares as a function of the heliocentric angle. The average flare-associated increase in $X$-ray flux was calculated for each degree of heliocentric angle and is shown by the blue dots, while the red histogram bars indicate the number of flares analyzed. The appearance of many flares in the range 87$^{\circ}$–90$^{\circ}$ is due to projection effects, meaning that flares originating just beyond the solar limb may have been included in the analysis \cite{Hannah_etal_2011, Hayes_etal_2021}. As shown in Figure~\ref{Center_to_limb_Xray}, the intensity of $X$-ray radiation does not vary significantly with the flare location on the solar disk. Therefore, we further normalize the EUV flux by the $X$-ray flux to reduce the observed dispersion.

\begin{figure} [h]
        \noindent\includegraphics[width=420pt]{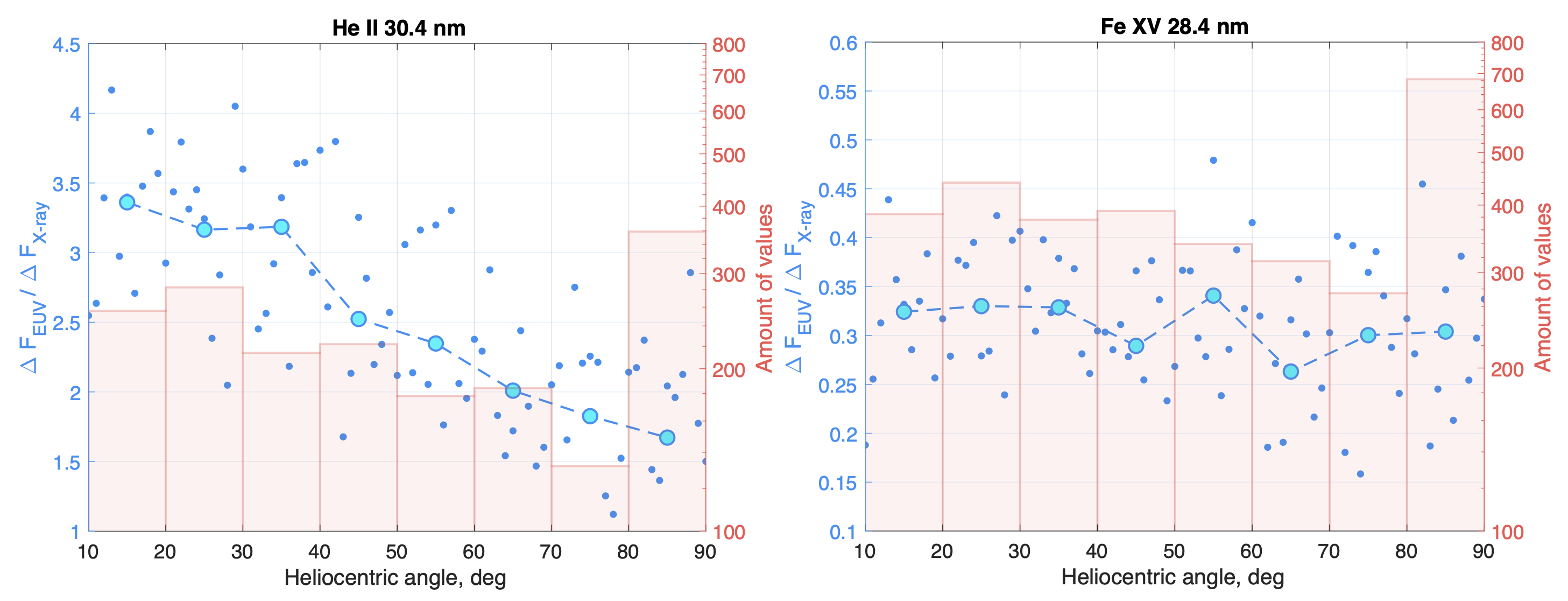}
        \centering
        \caption{Dependence of the flux increase in 30.4 nm (left) and 28.4 nm (right) lines relative to the $X$-ray flux ($\Delta F_{EUV}$/$\Delta F_{Xray}$) on the heliocentric angle. The blue dots represent the average values of $\Delta F_{EUV}$/$\Delta F_{Xray}$ for each degree of heliocentric angle; the cyan circles show the average values for 10-degree intervals; and the red histogram bars indicate the number of flares.}
    \label{Center_to_limb_EUV}
    \end{figure}

Figure~\ref{Center_to_limb_EUV} displays the flux increase ($\Delta F_{EUV}$) in the He~II 30.4 nm (left) and Fe~XV 28.4 nm (right) lines, normalized to $\Delta F_{Xray}$ (i.e., the flare class), as measured by the EVE (Extreme Ultraviolet Variability Experiment) instrument aboard the SDO satellite over several years. The blue dots represent the average values of $\Delta F_{EUV}$/$\Delta F_{Xray}$ for each heliocentric angle, while the cyan circles show the averages for 10-degree intervals. The red histogram bars indicate the number of flares analyzed. The He~II 30.4 nm line is optically thick and during solar flares the bulk of this emission is though to originate from the chromospheric footpoints of the flare loop \cite{Fletcher_etal_2011}. Consequently, the emission in this line is more strongly absorbed for flares near the solar limb than for those on the solar disk, as it passes through a greater thickness of the solar atmosphere. This heliocentric angle dependence is illustrated in the left panel of Figure~\ref{Center_to_limb_EUV}. In contrast, the Fe~XV 28.4 nm emission is formed in the solar corona, a low-density region where photon absorption and scattering are minimal. As a result, similar to $X$-rays, the emission is optically thin and independent of the flare's location on the solar disk, as clearly demonstrated in the right panel of Figure~\ref{Center_to_limb_EUV}.

Since the emissions at Fe~XV 28.4 nm and He~II 30.4 nm characterize the late and impulsive phases of a solar flare, respectively, the ratio of $Ne$ increase during the late phase relative to that during the impulsive phase is expected to be greater for flares located closer to the solar limb. However, this effect can only be reliably observed when analyzing a sufficient number of events because each solar flare has a unique spectrum.

\section{Experimental Data and $TEC$ Calculation}
\subsection{Solar Data}

As discussed in the Introduction, one of the most geoeffective emissions during the impulsive phase of a solar flare is He~II 30.4 nm, which ionizes the main neutral components of the ionospheric $F$ region ($O$, $N_2$, $O_2$), leading to a significant increase in electron concentration. In contrast, the primary emissions of the late phase of a solar flare are the iron lines at wavelengths of 28.4 nm and 33.5 nm. To analyze and compare the $TEC$ response to the impulsive and late phases, continuous long-term measurements of solar radiation flux at 30.4 nm, 28.4 nm, and 33.5 nm were required. For this purpose, we utilized data from the SDO/EVE MEGS-A (Multiple EUV Grating Spectrograph) instrument, which provided flux measurements of solar spectrum emissions in the 6–37 nm range with a 10-second cadence from April 2010 to May 2014. Additionally, since February 2017, we have access to measurements from the GOES-R/EUVS \cite{Eparvier_etal_2009, Snow_etal_2009} instrument, which observes several solar spectrum emissions, including He~II 30.4 nm and Fe~XV 28.4 nm, with a 1-second cadence. For this study, however, we employed calibrated Level 2 EUVS data with a 1-minute cadence.

Thus, we considered all X-class flares that occurred between April 30, 2010, and May 27, 2014, and between February 7, 2017, and December 31, 2024, to identify those exhibiting late phases. According to the definition of the EUV late phase introduced by \citeA{Woods_etal_2012}, this phase is characterized by a second peak in the emission of warm coronal lines, where the set of posteruptive coronal loops is higher than the first set of postflare loops; and the second peak of radiation is comparable in magnitude to the first. Consequently, we selected only the flares in which the second peak of Fe~XV emission reached at least 80\% of the intensity of the first peak in the same emission. Besides, to estimate the increase in total electron content, we need to subtract the trends associated with the GPS satellite flyby trajectory \cite{Bekker_etal_2024}. Since it is technically challenging to properly detrend $TEC$ curves during extended periods, we did not consider flares with late phases lasting more than 1.5 hours. We also excluded flares that exhibited additional emissions from hot coronal or cold chromospheric lines during the late phase, as these would interfere with the evaluation of the EUV late phase’s effect. Using these criteria, we identified 15 X-class flares with prominent late phases over the entire study period. Of these, 14 flares had similar power, ranging from X1.0 to X3.0, while one flare (X8.79 on May 14, 2024) was significantly more intense and thus excluded from the final sample. The final list of 14 flares, along with their key characteristics, including flare class, $X$-ray peak time, peak time of Fe~XV emission, late phase duration, flare location on the solar disk, and measurement instrument, can be found in Table~\ref{T1}.

\begin{table}[h]
    \caption{Characteristics of X-Class Solar Flares with a Late Phase}
    \centering
        \begin{tabular}{|p{0.25cm}|p{2.1cm}|p{0.9cm}|p{1cm}|p{1.1cm}|p{1.8cm}|p{2.3cm}|p{1.8cm}|}
        \hline
        \centering N & \centering Date (dd.mm.yyyy) & \centering Flare Class & \centering $X$-ray Peak Time & \centering Fe~XV Peak Time & \centering Late Phase Duration (min) & \centering Flare Location / Heliocentric Angle & Satellite / Instrument \\
        \hline
        \hline
          \centering 1 & \centering 22.09.2011 & \centering X2.14 & \centering 11:01 & \centering 12:22 & \centering 42 & \centering N09E89 / 89$^{\circ}$ & \multirow{5}{1.8cm}{\centering SDO / MEGS-A} \\
          \centering 2 & \centering 03.11.2011 & \centering X2.85 & \centering 20:27 & \centering 21:07 & \centering 24 & \centering N21E64 / 64$^{\circ}$ &   \\
          \centering 3 & \centering 23.10.2012 & \centering X2.37 & \centering 03:17 & \centering 04:45 & \centering 72 & \centering S13E58 / 58$^{\circ}$ &   \\
          \centering 4 & \centering 01.01.2014 & \centering X1.42 & \centering 18:52 & \centering 21:36 & \centering 75 & \centering S16W45 / 46$^{\circ}$ &   \\
          \centering 5 & \centering 07.01.2014 & \centering X1.04 & \centering 10:13 & \centering 11:13 & \centering 36 & \centering S13E11 / 17$^{\circ}$ &   \\
     \hline
          \centering 6 & \centering 08.09.2017 & \centering X1.16 & \centering 07:49 & \centering 08:41 & \centering 24 & \centering S10W57 / 57$^{\circ}$ & \multirow{9}{1.8cm}{\centering GOES-R / EUVS} \\
          \centering 7 & \centering 03.05.2022 & \centering X1.10 & \centering 13:25 & \centering 15:10 & \centering 42 & \centering S30E88 / 88$^{\circ}$ &   \\
          \centering 8 & \centering 06.01.2023 & \centering X1.22 & \centering 00:57 & \centering 02:39 & \centering 66 & \centering S20E82 / 82$^{\circ}$ &   \\
          \centering 9 & \centering 02.07.2023 & \centering X1.00 & \centering 23:14 & \centering 00:46 & \centering 45 & \centering N16W60 / 60$^{\circ}$ &   \\
          \centering 10 & \centering 22.02.2024 & \centering X1.70 & \centering 06:32 & \centering 07:07 & \centering 45 & \centering N17E35 / 37$^{\circ}$ &   \\
          \centering 11 & \centering 05.05.2024 & \centering X1.20 & \centering 11:54 & \centering 14:23 & \centering 54 & \centering N25W23 / 32$^{\circ}$ &   \\
          \centering 12 & \centering 11.05.2024 & \centering X1.54 & \centering 11:44 & \centering 12:33 & \centering 12 & \centering S19W60 / 60$^{\circ}$ &   \\
          \centering 13 & \centering 29.05.2024 & \centering X1.45 & \centering 14:37 & \centering 16:49 & \centering 30 & \centering S19E64 / 64$^{\circ}$ &   \\
          \centering 14 & \centering 10.06.2024 & \centering X1.55 & \centering 11:08 & \centering 11:51 & \centering 33 & \centering S17W89 / 89$^{\circ}$ &   \\
           \hline
        \end{tabular}
        \label{T1}
\end{table}

\begin{figure} [h]
        \noindent\includegraphics[width=420pt]{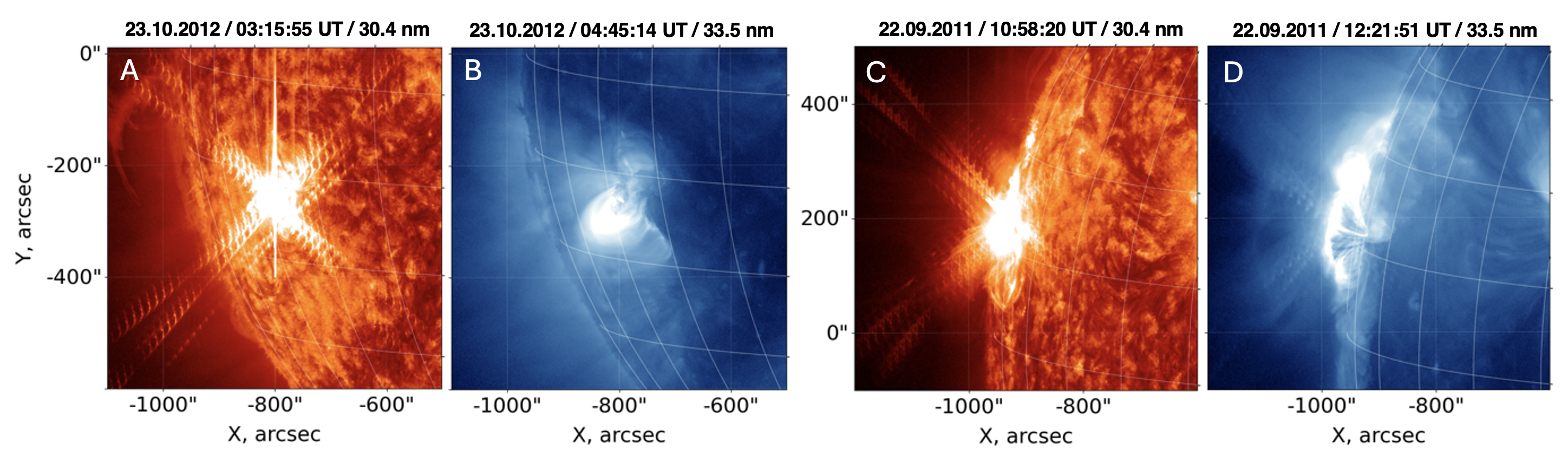}
        \centering
        \caption{SDO/AIA images of He~II 30.4 nm (a, c; impulsive phase) and Fe~XVI 33.5 nm (b, d; late phase) emissions for the disk flare on October 23, 2012 (a, b) and the limb flare on September 22, 2011 (c, d).}
    \label{AIA}
    \end{figure}

Below, this section presents a detailed comparative analysis of the results for two flares from the list: one that occurred on the solar disk (October 23, 2012) and the other on the solar limb (September 22, 2011). Figure~\ref{AIA} shows images of the solar disk captured by the SDO/AIA (Atmospheric Imaging Assembly; \cite{Lemen_etal_2012}) instrument during both the impulsive (a, c) and late (b, d) phases of these flares.

As seen in Figure 3, the footpoints of the limb flare are nearly invisible (c, d), suggesting that the measured radiation flux in the He~II line should be significantly weaker compared to a flare with the same initial spectrum located closer to the center of the solar disk. In contrast, emissions from the Fe~XV and Fe~XVI lines originate from coronal loops, which are clearly visible in Figure~\ref{AIA} for both the disk flare (b) and the limb flare (d). Since these emissions are less affected by attenuation, their flux should remain relatively constant, regardless of the flare's position on the solar disk.

\begin{figure} [h]
        \noindent\includegraphics[width=420pt]{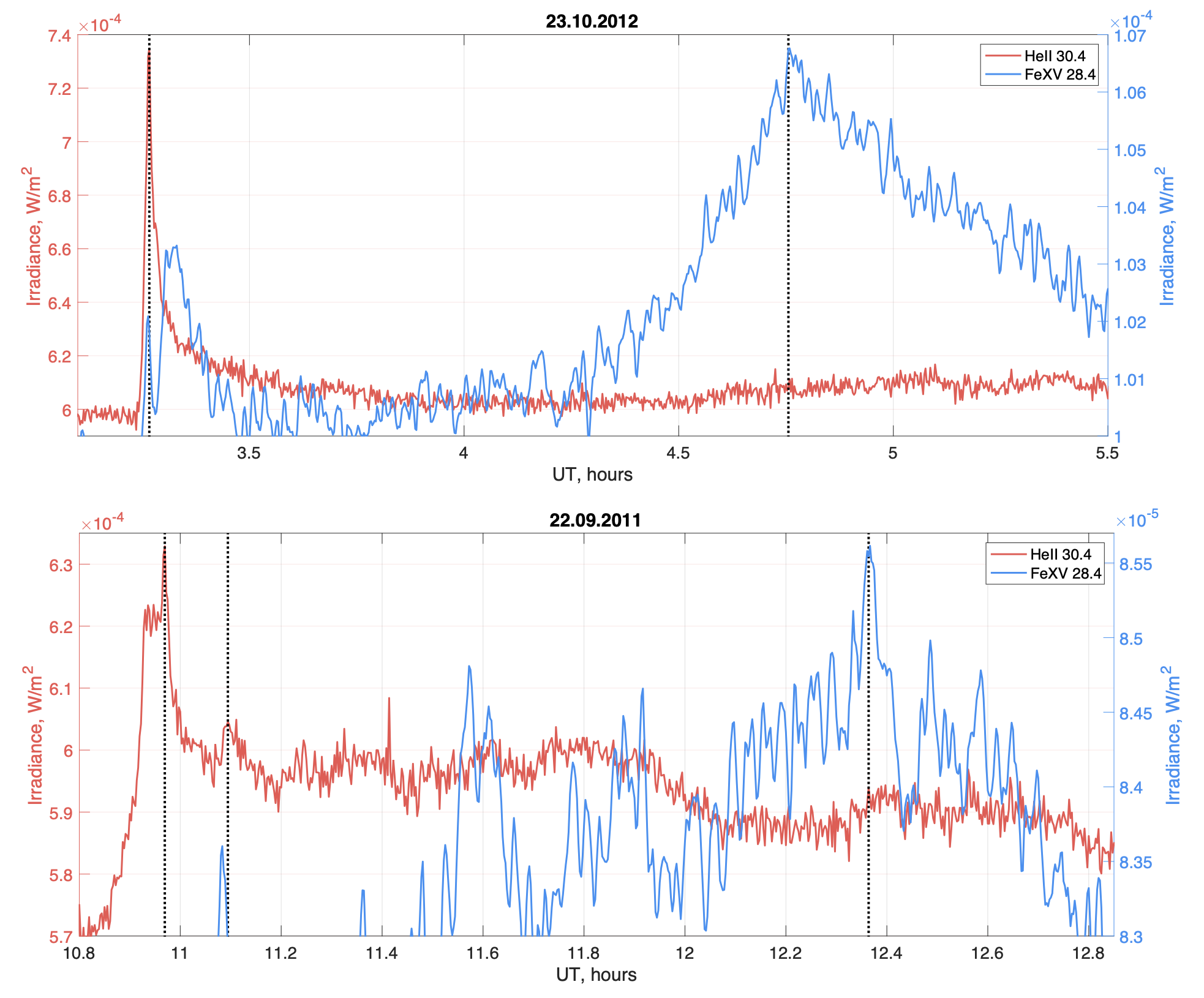}
        \centering
        \caption{Solar flare lightcurves in He~II 30.4 nm and Fe~XV 28.4 nm emissions measured by SDO/EVE during disk flare on October 23, 2012 (top panel) and limb flare on September 22, 2011 (bottom panel). Vertical dotted lines correspond to the emission peak times: 03:16 UT (He~II 30.4 nm) and 04:45 UT (Fe~XV 28.4 nm) on 23.10.2012; 10:58 UT, 11:06 UT (He~II 30.4 nm) and 12:22 UT (Fe~XV 28.4 nm) on 22.09.2011.}
    \label{Irradiance}
    \end{figure}

Figure~\ref{Irradiance} displays the flux dynamics in the He~II (red) and Fe~XV (blue) lines, as measured by the SDO/EVE instrument, for both the impulsive and late phases of the same flares. The blue curves are smoothed with a 1-minute time step. The graphs clearly highlight the impulsive and late phases, whose effects on the $TEC$ increase will be discussed in Section 3.2. Vertical dotted lines mark the moments when the emissions peaked. As expected, the flux increase ratio $\Delta F$(Fe~XV)/$\Delta F$(He~II) during the limb flare (September 22, 2011) is higher than for the disk flare (October 23, 2012). However, it is important to note again that while this trend is expected to hold statistically, it may not apply to individual flares, as each flare has a unique spectrum influenced by factors beyond just the optical thickness of the lines, including other physical mechanisms.

Since Fe~XVI irradiance measurements were unavailable for 9 of the 14 flares listed in Table~\ref{T1} (as we only had images from SDO/AIA for this line), the Fe~XV line is used to analyze the late phase of the flares in the following sections.

\subsection{Ionospheric Data}

For the numerical estimation of the $TEC$ increase during different phases of solar flares, we utilized measurements from the SOPAC GNSS network, which consists of thousands of GPS stations worldwide. The total electron content in the ionosphere was calculated by analyzing signal delays between GPS satellites and ground stations, using the methodology described by \citeA{Leonovich_etal_2002} and \citeA{Afraimovich_2000}.

\begin{figure} [h]
        \noindent\includegraphics[width=420pt]{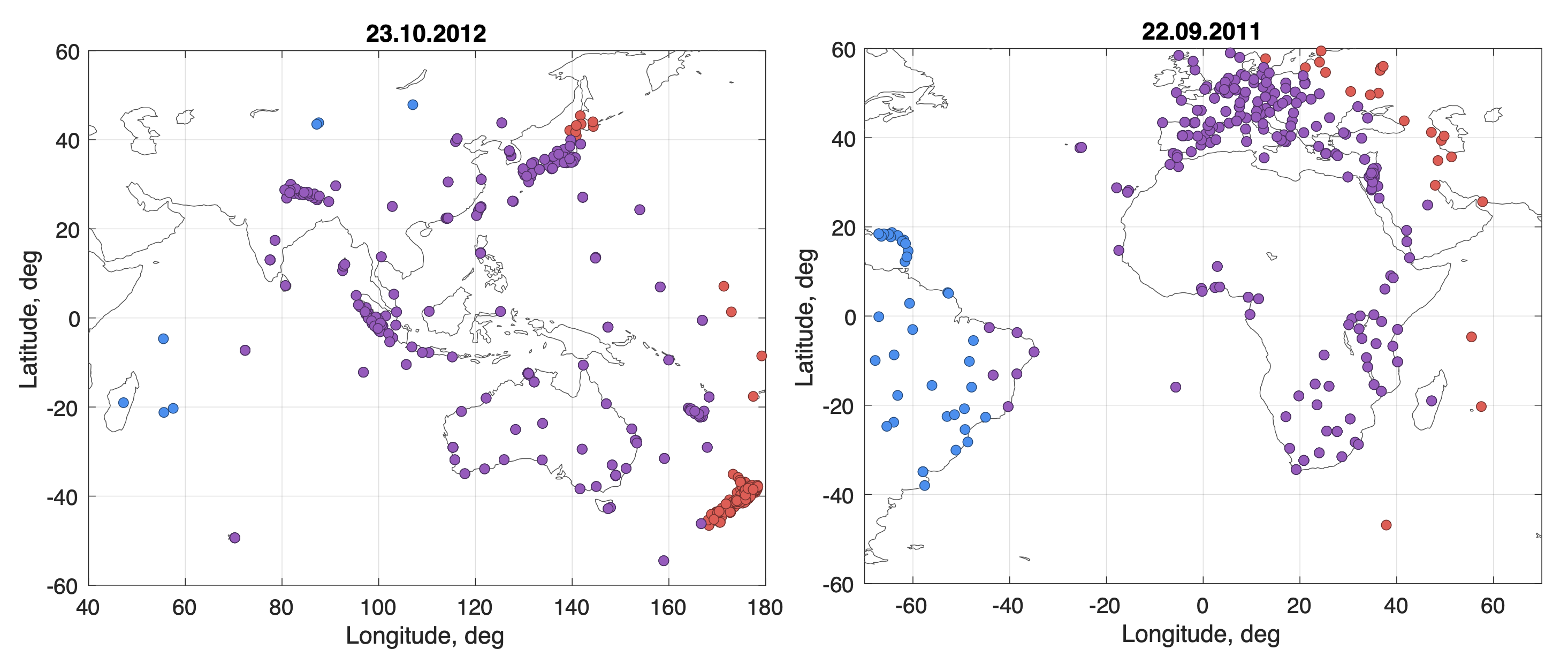}
        \centering
        \caption{Map of GNSS stations used for $TEC$ estimation during the flare of October 23, 2012 (left) and the flare of September 22, 2011 (right). Stations where the solar elevation angle was above 30$^{\circ}$ only during the impulsive phase are marked in red; those where the solar elevation angle was above 30$^{\circ}$ only during the late phase are marked in blue; and those where the solar elevation angle was above 30$^{\circ}$ during both phases are marked in purple.}
    \label{Map}
    \end{figure}

In this study, we selected stations located within the latitudinal range of –60$^{\circ}$ to +60$^{\circ}$, with a solar elevation angle greater than 30$^{\circ}$ during the selected flare events. Figure~\ref{Map} shows the locations of the stations for the flares on October 23, 2012 (left) and September 22, 2011 (right). Stations where the elevation angle was above 30$^{\circ}$ only during the impulsive phase are marked in red, those where it was above 30$^{\circ}$ only during the late phase are marked in blue, and stations where the elevation angle was above 30$^{\circ}$ during both phases are marked in purple.

During the impulsive phase of the flare on October 23, 2012, data from 401 ground stations were processed, while during the late phase, data from 212 stations were used. For the flare on September 22, 2011, the corresponding values were 257 and 274 stations, respectively. On average, about 4–5 satellites are visible simultaneously from each station; however, only those with an elevation angle greater than 30$^{\circ}$ were included in the analysis to minimize errors in the $TEC$ calculations.

\begin{figure} [h]
        \noindent\includegraphics[width=420pt]{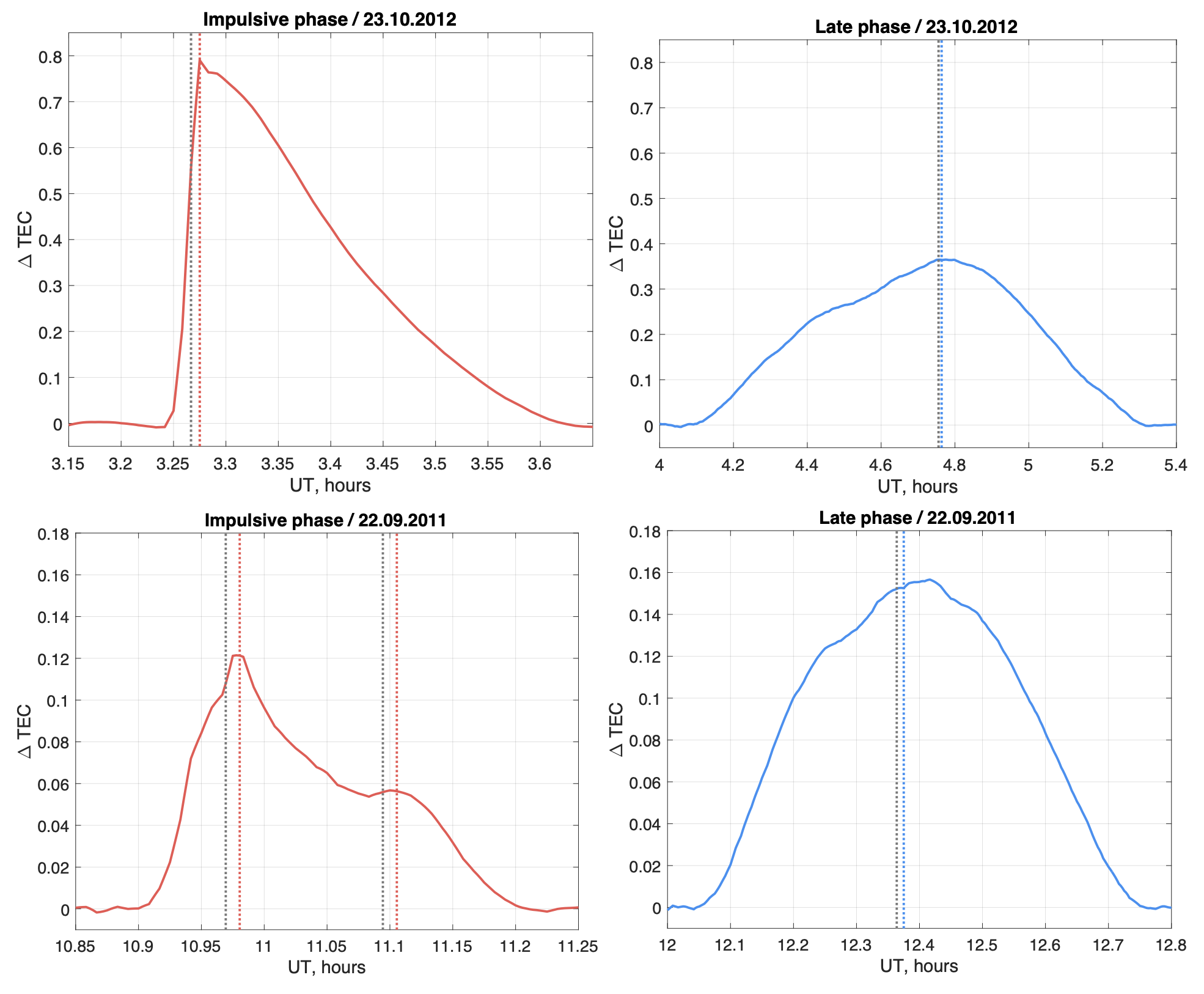}
        \centering
        \caption{$TEC$ increment during the impulsive (left panels) and late phases (right panels) of the disk flare on 2012 October 23 (top panels) and the limb flare on 2011 September 22 (bottom panels). Vertical grey lines correspond to the emission peaks, vertical colored lines correspond to the ionospheric response; time delay of 30 seconds for the flare on 2012 October 23; time delay of 40 seconds for the flare on 2011 September 22.}
    \label{TEC_dyn}
    \end{figure}

To calculate the $TEC$ increment ($\Delta TEC$) caused by different phases of the flare, we employed the detrending method described in \cite{Bekker_etal_2024,Bekker_2025}. During periods when increases were observed in the analyzed emissions, parabolic trends associated with satellite trajectories were subtracted from the relative $TEC$ measurements. These trends were approximated using third-order polynomials fitted over short time intervals immediately before and after the disturbance. The resulting average $\Delta TEC$ dynamics for the disk flare (October 23, 2012) and the limb flare (September 22, 2011) are shown in the upper and lower panels of Figure~\ref{TEC_dyn}, respectively. Vertical gray dotted lines indicate the times when the radiation fluxes reached their peaks (as shown in Figure~\ref{Irradiance}). Red dotted lines represent the ionospheric response times to He~II emission: for the flare on October 23, 2012, the ionospheric delay was approximately 30 seconds, while for the flare on September 22, 2011, it was about 40 seconds. Since 10-second measurements of $\Delta F$(He~II) and 15-second measurements of $\Delta TEC$ were used to calculate the delay, the accuracy of the response time estimate can be approximated as $\sqrt{(10/2)^2+(15/2)^2}\approx9$ seconds. Previous studies have reported $F$ region delays ranging from 20 seconds to 1 minute \cite{Bekker_etal_2024, OHare_etal_2025a, OHare_etal_2025b}, indicating that the values obtained here are consistent with earlier findings. These differences in response time are linked to the impulsiveness of the flare and the integral recombination rate in the ionosphere, as photochemical processes vary under different heliogeophysical conditions. The delays obtained for the impulsive phases were added to the times of the Fe~XV emission peaks to approximately determine the ionospheric response time to the late phases. These moments are indicated by vertical dotted blue lines in Figure~\ref{TEC_dyn}. After the peak of the limb flare late phase (bottom right panel of Figure~\ref{TEC_dyn}), a slight increase in the He~II flux occurred, causing the absolute maximum of $\Delta TEC$ to be reached approximately 3.5 minutes later. Nevertheless, the maximum $\Delta TEC$ value is nearly identical to the value at the point where the vertical blue line intersects the $\Delta TEC$ curve. Therefore, to calculate the increase in $Ne$ for the other events, we used the maximum values of the $\Delta TEC$ curves during both phases.

As shown in Figure~\ref{TEC_dyn}, the $TEC$ response to the late phase of the disk flare (top right panel) was $\sim$50\% of the response observed during the impulsive phase (top left panel). In contrast, the response to the late phase of the limb flare (bottom right panel) exceeded that of the impulsive phase (bottom left panel). This suggests that the center-to-limb variation effect is also evident in the ionospheric data, which is consistent with the patterns of chromospheric and coronal emissions discussed in Section 2. A similar calculations and analysis for all the flares listed in Table~\ref{T1} are presented in Section 4.

\section{Empirical Relationship Between $TEC$ Increase and Solar Irradiance During Different Phases of Solar Flares}

\begin{figure} [h]
        \noindent\includegraphics[width=420pt]{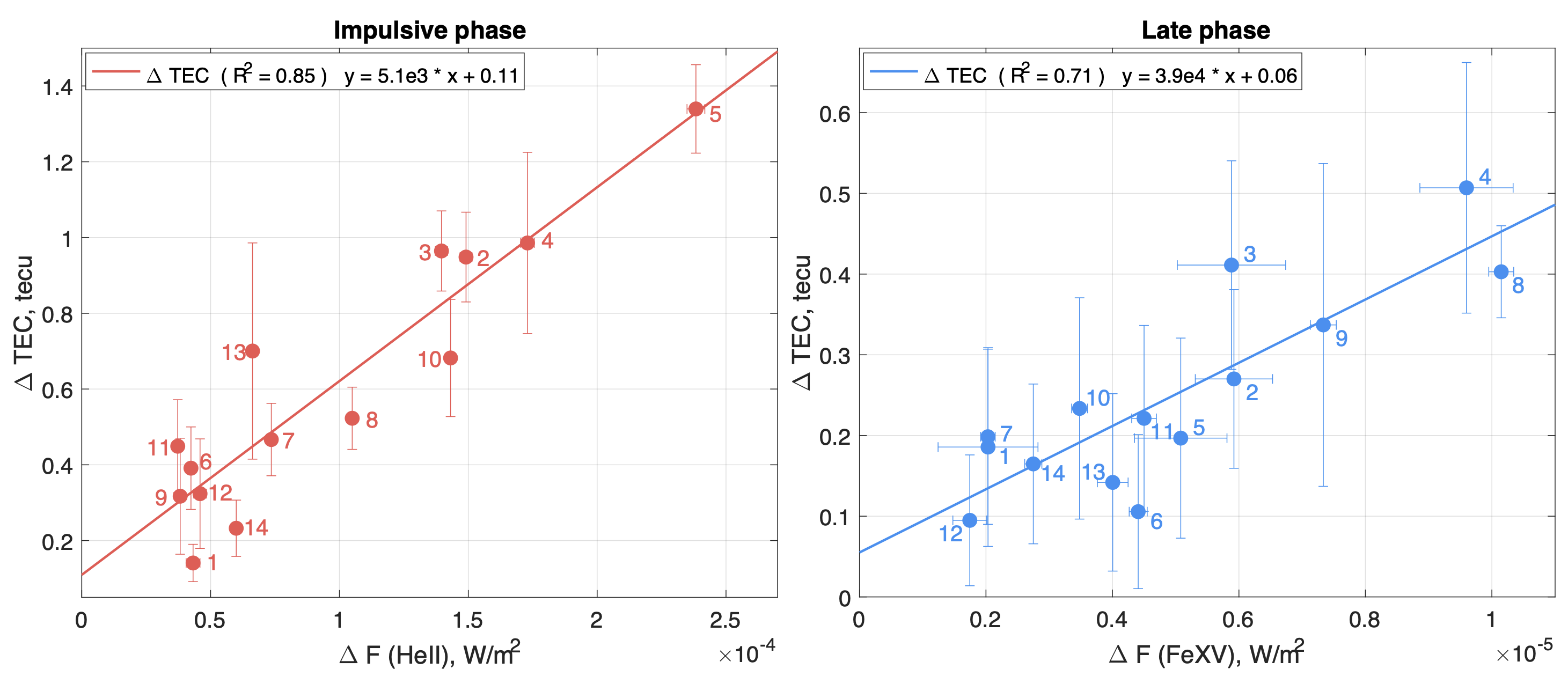}
        \centering
        \caption{Dependence of $\Delta TEC$ on $\Delta F$(He~II) during the impulsive phase (left) and on $\Delta F$(Fe~XV) during the late phase (right).}
 \label{dTEC}
 \end{figure}

Figure~\ref{dTEC} illustrates the relationship between the increase in total electron content during the impulsive (left) and late (right) phases of solar flares and the increase in flux in the He~II 30.4 nm (denoted as $\Delta F$(He~II)) and Fe~XV 28.4 nm (denoted as $\Delta F$(Fe~XV)) lines, respectively, based on data from 14 flares listed in Table~\ref{T1}. To account for the varying levels of ionospheric illumination at different measurement stations, the $\Delta TEC$ values are normalized by the square root of the sine of the average solar elevation angle \cite{Ieda_etal_2014}, as detailed in Table~\ref{T2}. Uncertainties in $\Delta F$ were determined by calculating the root mean square variations in flux during a quiet hour on the day of the flare for each line. Uncertainties in $\Delta TEC$ were estimated as the root mean square errors of the values obtained from different GPS stations and satellites for the given flare.

As shown in both panels, there is a clear increase in electron concentration with the increase in radiation flux in the respective lines. These dependencies are best described by a linear trend, which is expected given that, during a rapid increase in ionization, $\Delta Ne$ should be proportional to the ionization rate, which is itself linearly dependent on the flux of geoeffective radiation. For both phases, approximation lines were fitted to the data, and the coefficient of determination ($R^2$) was calculated. The coefficient of determination for the impulsive phase is $R^2 = 0.85$, while for the late phase, it is $R^2 = 0.71$. The lower $R^2$ value for the late phase can be attributed to larger errors in $\Delta TEC$ calculations, as this phase is characterized by a more gradual increase in flux, making it more challenging to accurately detrend the data using GNSS measurements. Despite this, both relationships are statistically significant and can be used to estimate $TEC$ increases during different flare phases, provided that flux dynamics data for the respective solar emissions are available. Additionally, we assessed the correlation between $\Delta TEC$ and the increment in Fe~XVI 33.5 nm flux during the late phase using data from SDO/AIA images. The $R^2$ value for this emission was 0.67, which is slightly lower than that for Fe~XV. This result is expected, as it clearly demonstrates a correlation; however, Fe~XV is known to be more geoeffective than Fe~XVI \cite{Bekker_etal_2024}, so the electron concentration in the ionosphere should exhibit a stronger correlation with the 28.4 nm flux.

It is worth noting that our result for the impulsive phase is in excellent agreement with previous studies reporting a strong correlation between $\Delta TEC$ and variations in the 26–34 nm EUV flux. \citeA{Sreeraj_etal_2025} analyzed $TEC$ enhancements during 49 X-class flares in Solar Cycle 24, obtaining a correlation coefficient of R = 0.86. Similarly, \citeA{Le_etal_2013}, based on 167 X-class flares in Solar Cycle 23, reported R = 0.91 for the same wavelength range.

The scatter in the left panel of Figure~\ref{dTEC} can be attributed to the fact that the He~II line is not the sole source of ionization in the ionosphere during the main phase of a solar flare. Even if we assume that the entire EUV spectrum increases proportionally, this assumption does not hold for $X$-ray radiation, which behaves differently \cite{Tsurutani_etal_2005}. For instance, the 5th flare listed in Table~\ref{T1}, although nearly the weakest in terms of $X$-ray power, exhibits the largest flux increase in the He~II 30.4 nm line (Figure~\ref{dTEC}, left panel). It is well known that the contribution of $X$-ray radiation, which ionizes the $D$ region, to the overall ionization in the ionosphere varies for each flare's spectrum and can be significant. Using the empirical dependence of the $D$ region's contribution ($\Delta TEC_D$) to the total ionospheric ionization, as derived in \cite{Bekker_Ryakhovsky_2024}, along with results from the FISM2 model (Flare Irradiance Spectral Model; \cite{Chamberlin_etal_2020}), we found that the contribution of the lower ionosphere for the flares analyzed in this study ranges from 8\% to 25\%. By calculating the combined electron content increase in the $E$ and $F$ regions ($\Delta TEC_{EF}$ = $\Delta TEC$ – $\Delta TEC_D$), which are primarily ionized by EUV radiation, we derived an updated dependence on the He~II flux. The coefficient of determination for this updated relationship was only slightly higher by 0.01. Given that this correction was negligible for the set of flares under consideration, we decided to disregard it and continue analyzing the dependence of the total electron content ($\Delta TEC$) on the He~II flux during the impulsive phase of the flare.

\begin{table}[h]
\caption{Mean Values of Latitude and Solar Elevation Angle for the GNSS Stations Used to Calculate $\Delta TEC$ During Different Flares}
\centering
\begin{tabular}{|c|c|c|c|c|c|c|}
\hline
\multirow{2}{*}{N} & \multirowcell{2}{Date \\ (dd.mm.yyyy)} & \multirowcell{2}{Flare \\ Class} & \multicolumn{2}{c|}{Mean Absolute Latitude, deg} & \multicolumn{2}{c|}{Mean Elevation Angle, deg} \\
 & & & Impulsive Phase & Late Phase & Impulsive Phase & Late Phase \\
\hline
\hline
1  & 22.09.2011 & X2.14 & 37 & 33 & 48 & 43 \\
2  & 03.11.2011 & X2.85 & 34 & 34 & 37 & 37 \\
3  & 23.10.2012 & X2.37 & 30 & 22 & 45 & 52 \\
4  & 01.01.2014 & X1.42 & 28 & 32 & 38 & 47 \\
5  & 07.01.2014 & X1.04 & 19 & 19 & 58 & 52 \\
6  & 08.09.2017 & X1.16 & 33 & 35 & 48 & 48 \\
7  & 03.05.2022 & X1.10 & 26 & 28 & 40 & 47 \\
8  & 06.01.2023 & X1.22 & 33 & 37 & 60 & 53 \\
9  & 02.07.2023 & X1.00 & 38 & 39 & 50 & 46 \\
10 & 22.02.2024 & X1.70 & 25 & 27 & 47 & 45 \\
11 & 05.05.2024 & X1.20 & 30 & 29 & 47 & 47 \\
12 & 11.05.2024 & X1.54 & 31 & 29 & 53 & 41 \\
13 & 29.05.2024 & X1.45 & 29 & 36 & 47 & 49 \\
14 & 10.06.2024 & X1.55 & 34 & 34 & 57 & 49 \\
\hline
\end{tabular}
\label{T2}
\end{table}

To investigate the factors contributing to the deviations of points from the approximation line in the right panel of Figure~\ref{dTEC}, we analyzed the season, solar and magnetic activity, latitude, and solar elevation angle at the GPS stations used in the calculations. As previously mentioned, the obtained $\Delta TEC$ values were normalized by the elevation angle, which proved to be sufficient to account for illumination, as a result, further analysis showed no significant correlation with the behavior of the points in Figure~\ref{dTEC}. The correlation coefficients for season, the solar index $F10.7$, and the magnetic index $Ap$ were 0.02, –0.26, and –0.27, respectively, which is insufficient to confidently assert an influence of these parameters. However, the latitudinal influence on $TEC$ was more pronounced.

\begin{figure} [h]
        \noindent\includegraphics[width=420pt]{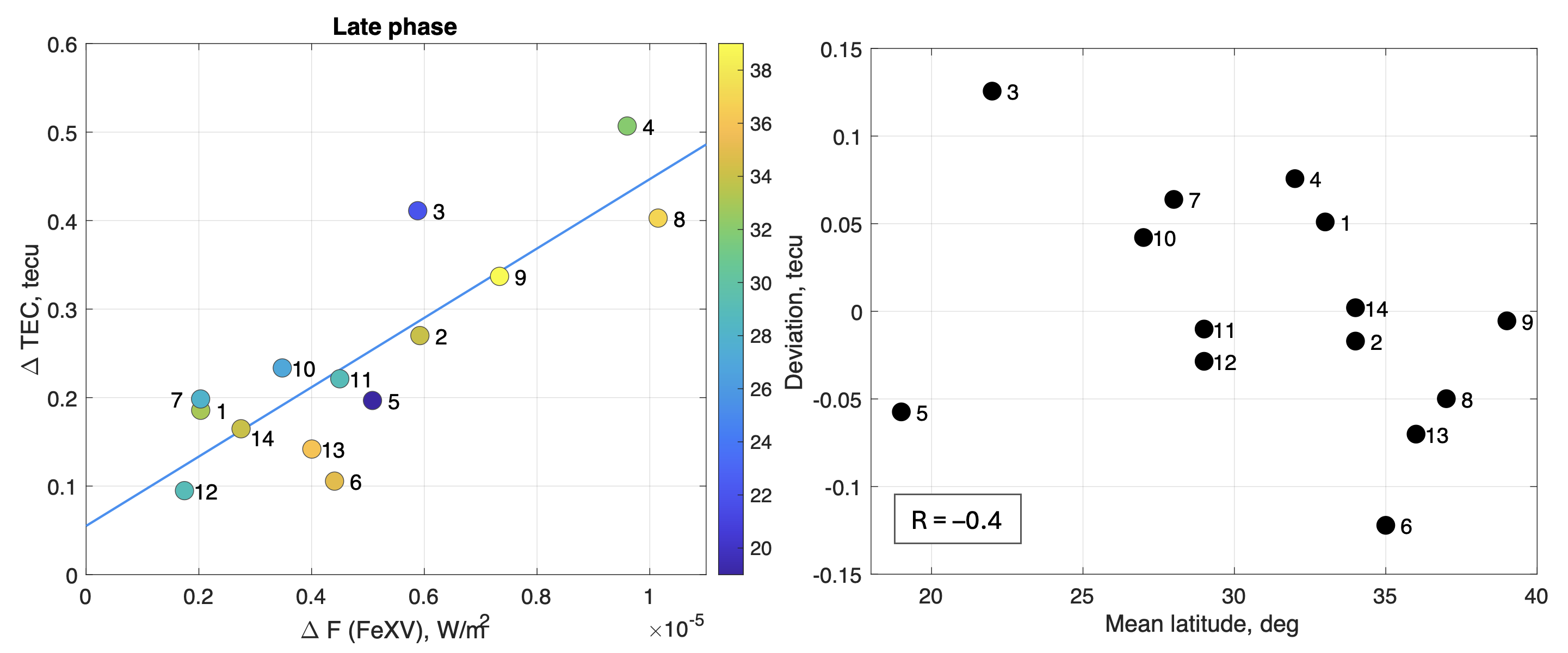}
        \centering
        \caption{Left panel: Dependence of $\Delta TEC$ on $\Delta F$(Fe~XV) during the late phase. The color scale indicates the average latitude of the GPS stations used for the $\Delta TEC$ calculation. Right panel: Latitudinal dependence of the deviation of $\Delta TEC$ from the fitting curve.}
    \label{Latitude}
    \end{figure}

In the left panel of Figure~\ref{Latitude}, we show the same dependence of $\Delta TEC$ on $\Delta F$(Fe~XV) as in the right panel of Figure~\ref{dTEC}, but with a color scale representing the average latitude of the GPS stations used for each selected flare. As observed, values below the approximation curve tend to correspond to higher latitudes. This behavior aligns with expectations, as the $TEC$ increase due to solar flares is generally greater at equatorial latitudes compared to mid-latitudes \cite{Yasyukevich_etal_2018}. The right panel of Figure~\ref{Latitude} illustrates the latitudinal dependence of the deviations of $\Delta TEC$ values from the approximation curve. A negative correlation between the parameters is visible, with the corresponding correlation coefficient of R = –0.4 indicating a moderate negative relationship. Excluding the 5th flare, which had an average latitude of 19$^{\circ}$, strengthens the correlation significantly (R = –0.7). A similar latitudinal effect is observed during the impulsive phase of the flares, though it is less pronounced. 

It is important to note that the variation in latitudes likely contributed significantly to the error values shown in the right panel of Figure~\ref{dTEC}. For instance, the widest latitude spread occurred during the late phase of the 9th flare, which demonstrates the largest error bar. 

Analyzing a larger number of events would make it possible to account for the influence of latitude and other parameters, thereby reducing the scatter in the results and deriving functional relationships for different heliogeophysical conditions.

As shown in Figure~\ref{dTEC}, the increase in $\Delta TEC_{late}$ during the late phase is proportional to the flux enhancement in the optically thin Fe~XV 28.4 nm line, while the increase in $\Delta TEC_{imp}$ during the impulsive phase is proportional to the flux enhancement in the optically thick He~II 30.4 nm line. Therefore, the ratio $\Delta TEC_{late}$/$\Delta TEC_{imp}$ is expected to increase with the heliocentric angle. As mentioned before, the larger the flare sample, the more pronounced this dependence should be.

\begin{figure} [h]
        \noindent\includegraphics[width=420pt]{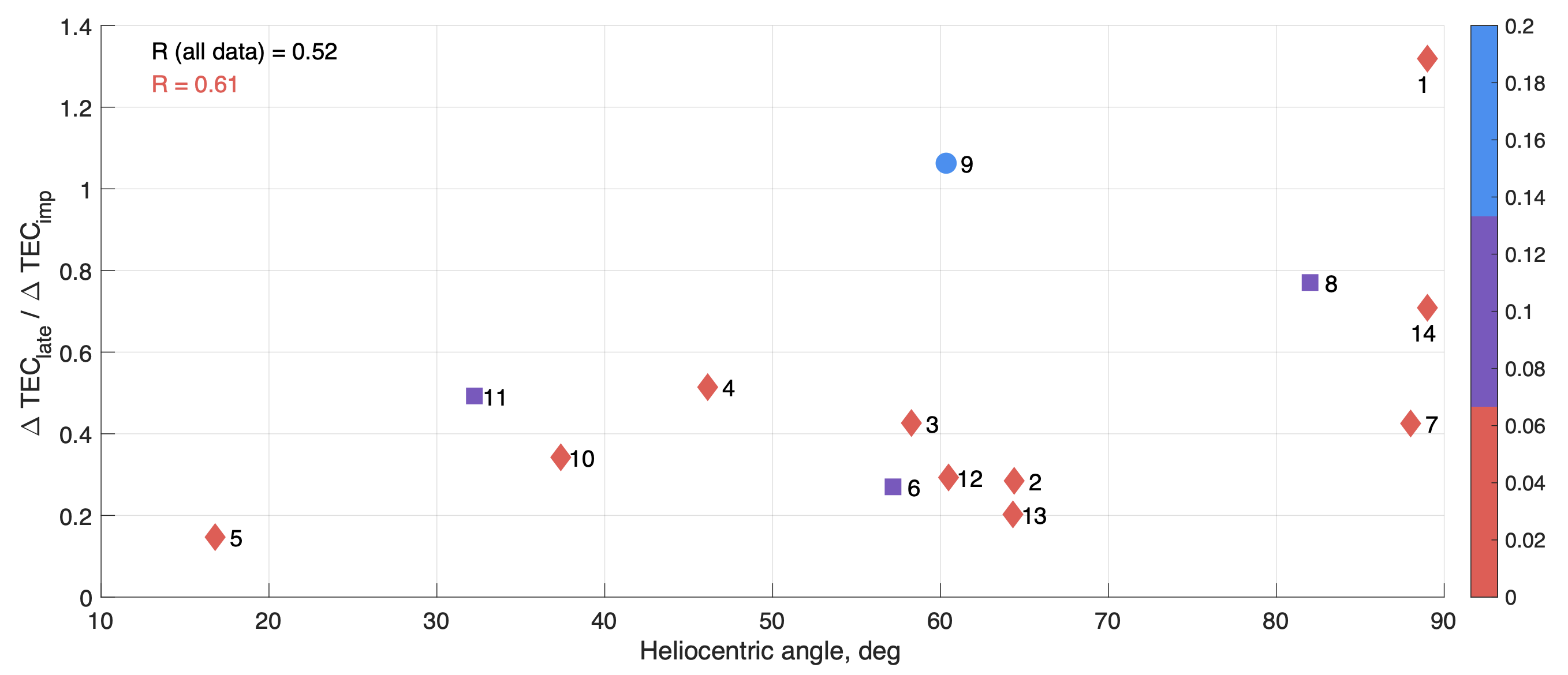}
        \centering
        \caption{Dependence of the $\Delta TEC_{late}$/$\Delta TEC_{imp}$ on the heliocentric angle. The color indicates the ratio of fluxes $\Delta F$(Fe~XV)/$\Delta F$(He~II).}
    \label{Location}
    \end{figure}

Figure~\ref{Location} shows the dependence of $\Delta TEC_{late}$/$\Delta TEC_{imp}$ on the heliocentric angle, derived from the 14 flares. Despite the limited number of events, we observe the expected increase in the ratio as the flare approaches the solar limb (R = 0.52). However, we can see significant scatter in the values for heliocentric angles around 60$^{\circ}$. Upon analyzing the spectra of these flares, we found that, although flares 2, 3, 9, 12, and 13 are located at roughly the same distance from the solar disk center, their spectra differ substantially. For example, the flux ratio $\Delta F$(Fe~XV)/$\Delta F$(He~II) for 9th flare is 0.192, while the ratio for flares 2, 3, 12, and 13 ranges from 0.038 to 0.060, which is approximately 3–5 times lower. To investigate the impact of the solar spectrum on this relationship, we added the flux ratio $\Delta F$(Fe~XV)/$\Delta F$(He~II) to the plot, using color to represent its value. For clarity, the values are divided into three equal ranges from 0 to the maximum value of 0.192 (9th flare). By excluding the blue and purple flares with ``non-standard'' spectra, the correlation coefficient increased to R = 0.61, making the dependence of the ratio of late to impulsive phase influences on $TEC$ more evident with respect to the heliocentric angle. This analysis demonstrates a connection between a solar flare’s location on the solar disk and its impact on variations in electron concentration in the Earth's ionosphere.

\section{Conclusion}

The Earth's ionosphere is an essential element of space weather, playing a vital role in the functioning of modern communication and navigation systems. Despite extensive research into how ionospheric parameters respond to various driving mechanisms, many questions about solar-terrestrial interactions remain unresolved, and the accuracy of ionospheric dynamics forecasting is still insufficient. One such unresolved issue involves the variations in ionospheric parameters induced by emissions from different phases of solar flares.

The impact of the late phase of a solar flare on the concentration of charged particles in the Earth's ionosphere has long been overlooked by the scientific community. The gradual increase in warm coronal radiation during the late phase does not induce the abrupt changes in electron concentration, as occurs during the impulsive phase. As a result, until 2024, its effects on the ionosphere had not been observed or analyzed. By utilizing extensive GNSS and solar observation data, along with advanced mathematical and statistical processing methods, we were able to quantitatively assess the influence of the late phase on the $TEC$ increase. Our analysis highlights the significant geoeffectiveness of warm coronal lines, whose contributions are often indistinguishable during the flare's main phase, when strong chromospheric EUV emissions and $X$-rays dominate.

This study presents an analysis of the ionospheric response to both the impulsive and late phases of 14 X-class solar flares, observed at varying distances from the center of the solar disk. It has been well established that the solar radiation spectrum changes significantly depending on the observation point. For instance, coronal emissions are largely unaffected by the heliocentric angle, while chromospheric lines can experience considerable attenuation when observed at the solar limb due to the thicker layers of the solar atmosphere they must pass through. This effect caused the flux of cold chromospheric lines (such as He~II 30.4 nm), which ionizes the ionosphere during the impulsive phase, to be statistically weaker for flares closer to the solar limb, while warm coronal emissions, which ionize the ionosphere during the late phase of the flare, are not subject to this attenuation. Through the analysis of a relatively small flare sample, we were able to observe this trend and demonstrate the importance of the late phase, which, under certain conditions, can exceed the impact of the more widely studied main phase.

Additionally, we derived empirical relationships between the increase in total electron content and the irradiance increase during both the impulsive and late phases of solar flares. These relationships provide a means for numerical estimation of the flare-induced effects on the Earth's ionosphere. Moreover, we identified and demonstrated the influence of GNSS station latitude on the deviation of $\Delta TEC$ values during the late phase. Accounting for this parameter in future analyses could reduce errors and enable the development of a more reliable empirical dependence of $\Delta TEC_{late}$ on $\Delta F$(Fe~XV).

This study has advanced our understanding of the effects of different phases of solar flares (considering their positions on the solar disk) on variations in the ionospheric total electron content and has outlined future directions for this research. First, the analysis demonstrated that both the solar zenith angle and the geographic latitude of GNSS stations must be accounted for. Second, incorporating data from additional instruments will enable the study of flares with more powerful and prolonged late phases, which last several hours and therefore cannot be adequately analyzed using the SOPAC GNSS network data. Third, continuous monitoring of geoeffective solar spectral emissions would expand the statistical sample of flares with late phases, leading to a more refined empirical patterns. Moreover, improving the temporal resolution of solar measurements would allow for a more detailed analysis of the ionospheric response, including the delay in $TEC$ increases, and enable the evaluation of the integral recombination rate in the Earth's ionosphere.

\section*{Open Research Section}

No new data were generated as part of this research. Data from the SOPAC network are available at \url{http://sopac-old.ucsd.edu/}. The FISM2 model data can be obtained under \url{https://lasp.colorado.edu/eve/data_access/eve_data/fism/}. The GOES-R and SDO data used in this study are available at \url{https://www.ngdc.noaa.gov/stp/satellite/goes-r.html} and \url{https://lasp.colorado.edu/eve/data_access/}, respectively. This work was (partly) carried out by using Hinode Flare Catalogue (\url{https://hinode.isee.nagoya-u.ac.jp/flare_catalogue/}), which is maintained by ISAS/JAXA and Institute for Space-Earth Environmental Research (ISEE), Nagoya University.

\section*{Acknowledgments}

S.Z.B. and R.O.M. would like to thank the European Office of Aerospace Research and Development (FA8655-22-1-7044-P00001) for supporting this research. R.O.M. would also like to acknowledge support from STFC New Applicant grant ST/W001144/1. This research was also supported by the International Space Science Institute (ISSI) in Bern, through ISSI International Team project \#24-618. I.A.R. would like to acknowledge support from Ministry of Science and Higher Education of Russian Federation (122032900175-6).

\bibliography{References}

\end{document}